\documentclass[journal,comsoc,10pt]{IEEEtran}
\usepackage[T1]{fontenc}% optional T1 font encoding
\usepackage{url}
\usepackage{graphicx}
\usepackage{color}
\ifCLASSINFOpdf
\else
\fi
\usepackage{cite}
\usepackage{amsmath}

\usepackage{color}
\usepackage[justification=centering]{caption}

\usepackage[cmintegrals]{newtxmath}
\usepackage{subcaption}
\usepackage{algorithm}  
\usepackage{algorithmic} 
\usepackage{amssymb}

\begin{document}
	\title{Elevation Angle-Dependent 3D Trajectory Design
for Aerial RIS-aided Communication}

	\author{	Yifan Liu,~\IEEEmembership{Student Member, IEEE},
		Bin Duo,~\IEEEmembership{Member, IEEE},
		Qingqing Wu,~\IEEEmembership{Senior Member, IEEE},
		Xiaojun Yuan,~\IEEEmembership{Senior Member, IEEE},
			Jun Li,~\IEEEmembership{Senior Member, IEEE},
		and Yonghui Li,~\IEEEmembership{Fellow, IEEE} \vspace{-0.5cm}
	
		\thanks{

Yifan Liu is with the College of Mechanical and Electrical Engineering, Chengdu University of Technology, Chengdu 610059, China. Bin Duo is with the College of Computer Science and Cyber Security, Chengdu University of Technology, Chengdu 610059, China. Qingqing Wu is with the State Key Laboratory of Internet of Things for Smart
City, University of
Macau, Macao 999078, China. Xiaojun Yuan is with the National Key Laboratory of Science and Technology
on Communications, University of Electronic Science and Technology of China,
Chengdu 611731, China. Jun Li is with the School of Electronic and Optical Engineering, Nanjing University of Science and Technology, Nanjing 210094, China. Yonghui Li is with the School of Electrical and Information Engineering, The University of Sydney, Sydney, NSW 2006, Australia.}}
	\maketitle
	\begin{abstract}
This paper investigates an aerial reconfigurable intelligent surface (RIS)-aided  communication system  under the probabilistic line-of-sight (LoS)  channel, where an unmanned aerial vehicle (UAV) equipped with an RIS is deployed to assist two ground nodes in their information exchange. An optimization problem with the objective of maximizing the minimum average achievable rate  is formulated to jointly design the communication scheduling, the RIS's phase shift, and the three-dimensional (3D) UAV  trajectory. To solve such a non-convex problem, we propose an efficient iterative algorithm to obtain its suboptimal  solution. Simulation results show that our proposed  design significantly outperforms the existing schemes and  provides new insights into the elevation angle and distance trade-off  for the UAV-borne RIS communication system.
	\end{abstract}
	

	% \fi
	\IEEEpeerreviewmaketitle

	\section{Introduction}
\textcolor{blue}{Besides terrestrial deployment, wireless networks are gradually evolving into air-ground integrated networks to achieve ubiquitous wireless connection and network capacity upgrades \cite{1,sha}. Recently, unmanned aerial vehicles (UAVs) have obtained substantial attention in wireless communication.} Thanks to their high mobility, low cost, and line-of-sight (LoS) transmission, UAVs  can further improve communication coverage, throughput, and average
secrecy rates \cite{2,3,DB}.

Recently, the reconfigurable intelligent surface (RIS) has attracted  considerable   attention  due to its low profile, low energy consumption,  and ability to overcome the  non-LoS (NLoS) transmission \cite{xi}. \textcolor{blue}{Typically, the RIS contains many reflective elements, each of which is manipulated to induce changes in the amplitude and phase shift of  incident signals to create favorable propagation environment.} The RIS becomes a promising technology for the future mobile communications. It can solve the pain points of fifth-generation (5G), such as high energy consumption and coverage voids. By optimizing the phase shift of all elements of the RIS, the signals  from different transmission paths can be precisely aligned at the desired receiver to increase the signal power  \cite{09}. 

For typical application scenarios of  UAV communications in urban areas, such as cargo delivery, traffic monitoring, and so on, their communication links  are often blocked by tall building, which leads to severe degradation of  channel quality. Fortunately, with its low power consumption and lightweight, the RIS can be installed at an appropriate location to reconfigure the propagation environment of air-ground links,  thereby improving communication performance. Several works have studied various RIS-assisted UAV communication systems. In general, these studies mainly fall into two categories, one for  terrestrial RIS\cite{D,liuyuanwei,glnz,8} and the other for aerial RIS\cite{A,B,zhangrui,fangfang,lala}. In particular, for the first category, the UAV trajectory and  the phase shift of the RIS mounted on building facades are jointly designed to optimize different objectives such as communication coverage \cite{D}, energy efficiency \cite{liuyuanwei}, confidentiality \cite{glnz}, and communication rates\cite{8}.

Taking advantage of  the UAV's  ability to fly freely in the three-dimensional (3D) space, the RIS can be mounted on the UAV. This allows the RIS to fly along with the UAV, which is more flexible in adjusting its 3D location than the conventional terrestrial RIS,  thus enhancing  communication services \cite{A,B}. An aerial RIS was deployed in \cite{zhangrui} to expand the coverage of communication services, where the worst-case signal-to-noise ratio (SNR) was improved by jointly optimizing the transmit beamforming, the RIS's placement, and the 3D passive beamforming. In \cite{fangfang}, multiple users were served by a base station (BS), and their received powers were significantly enhanced with the aid of an aerial RIS. To provide communication services for blocked users that locate far apart while preventing information leakage, the UAV equipped with one RIS was deployed in \cite{lala} to improve the security and energy efficiency. Above  works only took the ideal deterministic LoS channel (DLC) into account,  which has two limitations in practice. $i)$ The DLC model cannot fully capture the critical effects of the UAV location-dependent LoS and NLoS states in urban areas with typically high and dense buildings/trees\cite{7}; $ii)$ the DLC model  cannot accurately describe the elevation angle and distance trade-off, since the elevation angles between the UAV and ground nodes (GNs) are closely related to the 3D UAV trajectory \cite{6}. Therefore, it is intuitive that the UAV trajectory designed under the simplified DLC model  will undoubtedly cause significant performance loss  in practical urban environments. 
	
	Motivated by the above, this paper considers an aerial  RIS-aided  communication system, where the UAV-borne RIS assists in information exchange with two GNs. In particular, we adopt a more accurate probabilistic LoS channel  (PLC) model to characterize  the complex channel states of LoS and NLoS in an urban environment. To maximize the minimum average  achievable rate, we jointly optimize  the communication scheduling, the RIS’s phase shift, and the 3D UAV trajectory. The formulated problem is non-convex and difficult to solve, since  it contains intractable non-convex constraints, binary scheduling variables, and the complicated  achievable rate expression concerning UAV  trajectory variables. To tackle such challenges, we propose an efficient iterative algorithm to obtain a high-quality  solution. Simulation results show that our proposed joint design for the aerial RIS-aided communication system under the PLC model can significantly improve the  max-min rate compared to that under the conventional DLC model.  This is because the optimized UAV trajectory better balances the elevation angle and distance trade-off, resulting in the enhanced gain of the cascaded channel between the UAV and the GNs.
	
	\textcolor{blue}{Note that although the authors in \cite{ty} and \cite{Mah} also considered a UAV-borne RIS communication system under the PLC model, in this paper, both the system model and the proposed algorithm are different, which results in the fundamental trade-off between the elevation angle and distance for rate enhancement. Furthermore, in [17], the 3D UAV trajectory was designed for data collection under the PLC model; however, this paper considers a new UAV-borne RIS communication system and derives a new expression for the expected achievable rate function. }

\section{System Model and Problem Formulation}
 In this paper, we consider an  aerial  RIS-aided  communication system, where two GNs exchange information via a UAV-borne RIS\footnote{\textcolor{blue}{Note that the movement of the UAV changes the orientation of the RIS, which will cause the difference between the phase shift designed in this paper and the  required phase shift in practice, resulting in performance loss. However, it is reasonable to assume that the RIS can always remain stable in this paper, since we can employ a three-axis gimbal to ensure that the orientation of the RIS keeps unchangeable for the duration of the UAV flight \cite{8000387}.}} due to the blockage of dense buildings.   We characterize the position of the UAV and the two GNs via the 3D Cartesian coordinate system. It is assumed that the UAV flies  over a given duration $T$ to assist in reflecting signals via the RIS between the  GNs, whose locations are denoted by $\mathbf{w}_k =[x_k, y_k]$, $k\in\mathcal{K}=\left\lbrace 1, 2\right\rbrace $.  To ease the 3D UAV trajectory design,  $T$ is divided into $N$ time slots that are equal in length, i.e., $T= N\delta_{t}$, where $\delta_{t}$ is the length of each time slot. Thus, the  trajectory of the UAV can be approximated by a 3D sequence $\left\{\left(\mathbf{q}^{}[n], h[n]\right)\right\}, n \in \mathcal{N}=\left\lbrace 1,2,...,N\right\rbrace $, where   the discrete way-points $\mathbf{q}[n]= [x[n], y[n]]$ and $h[n]$ represent the horizontal and vertical locations, which satisfy the following constraints:
\begin{equation}
	\|\mathbf{q}[n+1]-\mathbf{q}[n]\|^{2} \leq \hat{\Omega}^{2}, \forall n,
\end{equation}
\begin{equation}
	\mathbf{q}[N+1]=\mathbf{q}_{F}, \mathbf{q}[1]=\mathbf{q}_{0},
\end{equation}
\begin{equation}
	|{h}[n+1]-{h}[n]|^{2} \leq\tilde{\Omega}^{2}, H_{\rm {min}}\leq h[n]\leq H_{\rm{max}}, \forall n, 
\end{equation}
where $\mathbf{q}_0$ and $\mathbf{q}_F$ denote the initial and final horizontal positions
of the UAV, respectively, $\hat{\Omega} = \hat{v}_{\rm {max}}\delta_{t}$ and $\tilde{\Omega} = \tilde{v}_{\rm {max}}\delta_{t}$ are the maximum horizontal and vertical distance that the UAV can reach in each time slot, respectively, and $\hat{v}_{\rm {max}}$ and $\tilde{v}_{\rm {max}}$  are the corresponding maximum horizontal and vertical flying speed of the UAV, respectively. Furthermore, $H_{\rm {min}}$ and $H_{{\rm {max}}}$ indicate the minimum and maximum altitude that the UAV can reach at any given time.

We assume that both GN 1 and GN 2 are equipped with an omni-directional single antenna. The RIS is equipped with a uniform planar array (UPA) of $M=M_x \times M_y$ ($M_x$ rows and $M_y$ columns) reflective elements, each of which  can be manipulated by an embedded development board, such as Raspberry Pi $4$, mounted on the UAV. Since the UAV and the RIS are assembled compactly, we assume that their 3D coordinates are identical, causing negligible performance loss due to high flying altitude of the UAV \cite{lala}. We denote $\Theta[n]=\operatorname{diag}\left\{e^{j \theta_{1,1}[n]}, e^{j \theta_{1,2}[n]}, \cdots, e^{j \theta_{M_x,M_y}[n]}\right\}$ as the
diagonal phase-shift matrix of the RIS in the $n$th time slot, where $\theta_{m_x,m_y}[n] \in[0,2 \pi)$ is the  phase shift  of $m_x$-th row and $m_y$-th column in the $n$th  time slot, where $m_x \in \mathcal{M}_x=\{1, \cdots, M_x\}$ and $m_y \in \mathcal{M}_y=\{1, \cdots, M_y\}.$ Assume that the phase shift of each element is continuously controllable, which should satisfy $|e^{j\theta_{m_x,m_y}[n]}|=1 $. 

To accurately represent the channel states in urban environments, we adopt the PLC model\cite{R1}, for which the ground-UAV channel can be represented by LoS or NLoS state. Thus, for GN $k$, the LoS probability in the $n$th time slot is given by
\begin{equation}
	P_{k}^{\mathrm{L}}[n]=\frac{1}{1+a e^{\left(-b\left[\psi_{k}[n]-a\right]\right)}},
\end{equation}
where \emph{a} > 0 and \emph{b} > 0 are constants specified by the actual environment, and

\begin{equation}
	\psi_{k}[n]=\frac{180}{\pi} \arctan \left(\frac{h[n]}{\left\|\mathbf{q}[n]-\mathbf{w}_{k}\right\|}\right)
\end{equation}
is the elevation angle from GN $k$ to the UAV in the $n$th  time slot. The relevant NLoS probability can then be  acquired as $P_{k}^{\mathrm{N}}[n]=1-P_{k}^{\mathrm{L}}[n]$.
  The channel gain between GN $k$ and the UAV conditioned on the LoS state in the $n$th time slot  can be expressed as
\begin{align*}
\mathbf{h}_{k}^{L}[n]=\tau[n]&\left[1,  \ldots, e^{-j \frac{2 \pi d}{\lambda}(M_x-1)  {\rm {sin}}\varphi_{k}[n]{\rm cos}\omega_k[n]}\right]^{T}\otimes\\
&\left[1,  \ldots, e^{-j \frac{2 \pi d}{\lambda}(M_y-1)  {\rm {sin}}\varphi_{k}[n]{\rm sin}\omega_k[n]}\right]^{T}, \tag{6}
\end{align*}
where $\tau[n]=\sqrt{\beta_{0} d_{k}^{-\alpha_{\mathrm{L}}}[n]}$,  $\beta_{0}$ is the path loss at the reference distance of $D_0=1$ meter (m), $d_k[n]=\sqrt{(\mathbf{q}[n]-\mathbf{w}_k)^2+h[n]^2}$ is the distance from GN $k$ to the UAV in the $n$th time slot, $\alpha_{\mathrm{L}}$ denotes the path loss exponent for the LoS  state. Furthermore, $d$ is the antenna separation,  $\lambda$ is the carrier wavelength,  $\varphi_{k}[n]$ and $\omega_k[n]$  represent the elevation and azimuth angles in the $n$th time slot, respectively. Furthermore, ${\rm {sin}}\varphi_{k}[n]{\rm cos}\omega_k[n]=\frac{x[n]-x_k}{d_k[n]}$ and ${\rm {sin}}\varphi_{k}[n]{\rm sin}\omega_k[n]=\frac{y[n]-x_k}{y_k[n]}$. \textcolor{blue}{ The channel gain between GN $k$ and the UAV conditioned on the NLoS state in the $n$th time slot  is given by
\begin{equation}
\mathbf{h}_{k}^{N}[n]=\zeta[n]\mathbf{\widetilde{h}{}}_{k}, \tag{7}
\end{equation}
where $\zeta[n]=\sqrt{\beta_{0} d_{k}^{-\alpha_{\mathrm{N}}}[n]}$,   $\alpha_{\mathrm{N}}$ denotes the path loss exponent for the NLoS  state, and $\mathbf{\widetilde{h}}_{k} \sim \mathcal{C N}(0,1)$  is the small-scale fading component
modeled by a circularly symmetric complex Gaussian (CSCG)
random variable.}

Assume that GN $k$ operates in the half-duplex mode, i.e., it can only receive or transmit  in each time slot. Thus we define a binary variable that indicates whether GN $k$ is scheduled to receive reflected signals from the UAV in the $n$th time slot or not, i.e.,  GN $k$ receives  signals from the other GN via the RIS if $\alpha_{k}[n]$ = 1,  and transmits otherwise. Assume that only one GN is allowed to transmit or receive signals to or from the UAV-borne RIS in the $n$th time slot, so we have the following scheduling constraints:
\begin{align*}
	&\sum_{k=1}^{2} \alpha_{k}[n] \leq 1, \forall n \in \mathcal{N},\tag{8}\\
	&\alpha_{k}[n] \in\{0,1\}, \forall k, n.\tag{9}
\end{align*}

 In a statistical sense, the expected achievable rate in bits/second/Hertz  (bps/Hz) from GN $\bar{k}$ through the UAV to GN ${k}$ in the $n$th time slot can be given by
	\begin{align*}
		\mathbb{E}\left[R_{k}[n]\right]=&P_{k}^{\mathrm{L}}[n] P_{\bar{k}}^{\mathrm{L}}[n] R_{k}^{\mathrm{LL}}[n]+P_{k}^{\mathrm{L}}[n] P_{\bar{k}}^{\mathrm{N}}[n] R_{k}^{\mathrm{LN}}[n] \\
		+&P_{k}^{\mathrm{N}}[n] P_{\bar{k}}^{\mathrm{L}}[n] R_{k}^{\mathrm{NL}}[n]+P_{k}^{\mathrm{N}}[n] P_{\bar{k}}^{\mathrm{N}}[n] R_{k}^{\mathrm{NN}}[n], \tag{10}
	\end{align*}
where $R_{k}^{{fo}}[n]=\log _{2}\left(1+{\gamma_{\bar{k}}\left|\left(\mathbf{h}_{k}^{f}[n]\right)^{H} \Theta[n] \mathbf{h}_{\bar{k}}^{o}[n]\right|^{2}}\right)$, $f,o\in\{L, N\}$
denote respectively the achievable rates at GN $k$ conditioned on the LoS and NLoS states of the air-ground channels.  Furthermore, $\gamma_{\bar{k}}=P_{\bar{k}}/\sigma^2$, where $P_{\bar{k}}$ is maximum transmit power of GN ${\bar{k}}$ and $\sigma^2$ is the variance of Gaussian noise at GN $k$. 

We aim to maximize the minimum average  achievable rate by jointly optimizing the communication scheduling $\mathbf{A}$, the horizontal UAV trajectory $\mathbf{Q}$, the vertical UAV trajectory $\mathbf{H}$, and the RIS's phase shift $\mathbf{\Theta}$ for  the entire $N$ time slot. Thus, the optimization problem can be formulated  as  
\begin{align*}
	\max _{\mathbf{A}, \mathbf{Q}, \mathbf{\Theta}, \mathbf{H}, \mathbf{\Psi_k}} &\quad\eta \tag{11a}\\
	\text { s.t. }&\frac{1}{N} \sum_{n=1}^{N} \alpha_{k}[n] 	\mathbb{E}\left[R_{k}[n]\right] \geq \eta, \forall k \in \mathcal{K}, \tag{11b}\\
	& |e^{j\theta_{m_x,m_y}[n]}|=1 , \quad \forall n, \tag{11c}\\
	&(1)-(3), (5), (8)-(9).
\end{align*}
Problem (11) is non-convex because the rate function $\bar{R}_{{k}}[n]$ in (11b) is  not jointly concave with respect to $\mathbf{A}$, $\mathbf{\Theta}$, $\mathbf{Q}$, and $\mathbf{H}$, the binary scheduling constraints in  (9) are non-convex, the phase shift constraints in (11c) are non-convex, and the elevation angle constraints in (5) are non-affine, which make it difficult to obtain the optimal solution. In the  following section, we propose an efficient  iteration algorithm to solve  problem (11).
	\section{Proposed Algorithm}
In this section, an alternating optimization method\footnote{The optimization algorithm can be executed on a ground control center. Specifically, the ground control center first executes the proposed algorithm and then transmits the optimized variables to the UAV through the control and non-payload communication (CNPC) link so that the UAV can perform the mission according to the designed trajectory  \cite{DUO}.} is proposed to obtain a
high-quality  solution to problem (11), where  $\mathbf{A}$, $\mathbf{\Theta}$, $\mathbf{Q}$, and $\mathbf{H}$ are iteratively optimized. Specifically,  the original problem is partitioned into three subproblems, each of which is solved  in an iterative manner  until the algorithm converges.
	\subsubsection{GNs'  Scheduling Optimization}
	With any given feasible  $\mathbf{Q}$, $\mathbf{\Theta}$, and $\mathbf{H}$, this subproblem can be expressed as
\begin{align*}
	\max _{\mathbf{A}, \eta} &\quad\eta  \tag{12a}\\
	\text { s.t. }& 0 \leq \alpha_{k}[n] \leq 1, \forall k, n,  \tag{12b}\\
	&(8),(11b) .
\end{align*}
Since problem (12) is a standard linear program, it can be solved efficiently by CVX \cite{cvx}.
\textcolor{blue}{\subsubsection{RIS's Phase Shift Design}
	With any given feasible  $\mathbf{A}$, $\mathbf{Q}$, and $\mathbf{H}$,  problem (11) can be rewritten as
	\begin{align*}
		&\max _{\Theta, \eta} \eta\tag{13} \\
		&\text { s.t. }(11 b)-(11 c). 
	\end{align*}
Since the constraints (11c) are unit modulus, it is difficult to solve problem (13). To overcome this difficulty, we employ the semidefinite relaxation
(SDR). To begin, let us perform the following transformation:
\begin{align*}
	\left(\mathbf{h}_{k}^{f}[n]\right)^{H} \Theta[n] \mathbf{h}_{\bar{k}}^{o}[n]=&\left(\mathbf{h}_{k}^{f}[n]\right)^{H}\operatorname{diag}(\mathbf{h}_{\bar{k}}^{o}[n])\boldsymbol{v}[n],\\
	& f,o \in \{L, N\},  \tag{14}
\end{align*}
where $	\boldsymbol{v}[n]=[e^{j \theta_{1,1}[n]}, e^{j \theta_{1,2}[n]}, \cdots, e^{j \theta_{M_{x}, M_{y}}[n]}]^T$.  According to (14), we can reinterpret the square term in the expected achievable rate as 
\begin{align*}
	&\left|\left(\mathbf{h}_{k}^{f}[n]\right)^{H} \Theta[n] \mathbf{h}_{\bar{k}}^{o}[n]\right|^2=\left(\mathbf{h}_{k}^{f}[n]\right)^{H}\operatorname{diag}(\mathbf{h}_{\bar{k}}^{o}[n])\boldsymbol{v}[n] \boldsymbol{v}^H[n]\\
	&\operatorname{diag}(\mathbf{h}_{\bar{k}}^{o}[n])^H\left(\mathbf{h}_{k}^{f}[n]\right)^{}=\operatorname{Tr}(\boldsymbol{V}[n]\mathbf{G}^{fo}[n]),\quad f,o \in \{L, N\}, \tag{15}
\end{align*}
where $\mathbf{G}^{fo}[n]=\operatorname{diag}(\mathbf{h}_{\bar{k}}^{o}[n])^H\left(\mathbf{h}_{k}^{f}[n]\right)^{}\left(\mathbf{h}_{k}^{f}[n]\right)^{H}\operatorname{diag}(\mathbf{h}_{\bar{k}}^{o}[n])$.
 Then, the expected achievable rate can be re-expressed as}
\textcolor{blue}{\begin{align*}
	\mathbb{E}\left[\widetilde{R}_{k}[n]\right]&=P_{k}^{\mathrm{L}}[n] P_{\bar{k}}^{\mathrm{L}}[n] \log_{2}\left(1+\operatorname{Tr}(\boldsymbol{V}[n]\mathbf{G}^{LL}[n])\right)\\&+P_{k}^{\mathrm{L}}[n] P_{\bar{k}}^{\mathrm{N}}[n] \log_{2}\left(1+\operatorname{Tr}(\boldsymbol{V}[n]\mathbf{G}^{LN}[n])\right) \\&+P_{k}^{\mathrm{N}}[n] P_{\bar{k}}^{\mathrm{L}}[n] \log_{2}\left(1+\operatorname{Tr}(\boldsymbol{V}[n]\mathbf{G}^{NL}[n])\right)\\&+P_{k}^{\mathrm{N}}[n] P_{\bar{k}}^{\mathrm{N}}[n] \log_{2}\left(1+\operatorname{Tr}(\boldsymbol{V}[n]\mathbf{G}^{NN}[n])\right). \tag{16}
\end{align*} 
\quad Since the expected achievable rate is a function of the newly introduced variable $\boldsymbol{V}[n]$, we can alleviate the difficulties associated with using $\boldsymbol{v}[n]$ as the optimization variable. In this respect,  $|\boldsymbol{v}_{m,m}[n]|=1$ can be rewritten as $\boldsymbol{V}[n]\succeq0$ and $\boldsymbol{V}_{m,m}[n]=1$ for the new optimization variable $\boldsymbol{V}[n]$. 
Thus, problem (13) can be rewritten as    
\begin{align*}
	\max _{\mathbf{\Theta}, \eta} &\quad \eta  \tag{17}\\
\text{s.t.}	& \frac{1}{N} \sum_{n=1}^{N} \alpha_{k}[n] 	\mathbb{E}\left[\widetilde{R}_{k}[n]\right] \geq \eta, \forall k \in \mathcal{K},\tag{17a}\\
&\boldsymbol{V}[n]\succeq0,
\tag{17b}\\
&\boldsymbol{V}_{m,m}[n]=1, m=1,2,...,M. \tag{17c}
\end{align*}	
As problem (17) is a semidefinite programming, it can be solved efficiently by CVX. However,  a rank-one
solution may not be obtained. Hence, we recover $\boldsymbol{v}[n]$ from $\boldsymbol{V}[n]$  using the Gaussian randomization method, which is similar to that in \cite{glnz} and thus omitted here for brevity.}
\textcolor{blue}{\subsubsection{UAV Horizontal Trajectory Optimization}
With any feasible $\mathbf{H}$ and $\mathbf{A}$ and $\mathbf{\Theta}$ obtained  by solving problem (12) and (17), respectively,  the UAV horizontal trajectory optimization problem can be written as
\begin{align*}
	&\max _{\mathbf{Q},\mathbf{\Psi_k}, \eta} \eta  \tag{18}\\
	&\quad\text { s.t. }(1 )-(2),(5),(11b).
\end{align*}
\quad To facilitate the solution to problem (18), we firstly rewrite Eq. (6) as $\boldsymbol{h}^L_k[n]=\tau[n]\boldsymbol{h}^{L^{'}}_k[n]$. Note that not only $\tau[n]$  but also $\boldsymbol{h}^{L^{'}}_k[n]$ is relevant to the  UAV trajectory. It is observed that $\boldsymbol{h}^{L^{'}}_k[n]$ is complex and non-linear with respect to the UAV trajectory variables, which makes the UAV trajectory design intractable. To handle such difficulty, we use the UAV trajectory of the $(l-1)$ th iteration to obtain an approximate $\boldsymbol{h}^{L^{'}}_k[n]$ in the $l$ th iteration \cite{glnz}.  Therefore, constraint (11b) can be re-expressed as
\begin{align*}
	& \frac{1}{N} \sum_{n=1}^{N} \alpha_{k}[n] \left(P_{k}^{\mathrm{L}}[n] P_{\bar{k}}^{\mathrm{L}}[n]\log _{2}\left(1+\xi_{k}^{L L}[n]d_{k}^{-\alpha_{\mathrm{L}}}[n] d_{\bar{k}}^{-\alpha_{\mathrm{L}}}[n]\right)\right.\\&\left.+P_{k}^{\mathrm{L}}[n] P_{\bar{k}}^{\mathrm{N}}[n]\log _{2}\left(1+\xi_{k}^{L N}[n]d_{k}^{-\alpha_{\mathrm{L}}}[n] d_{\bar{k}}^{-\alpha_{\mathrm{N}}}[n]\right)\right.\\&\left.
	+P_{k}^{\mathrm{N}}[n] P_{\bar{k}}^{\mathrm{L}}[n]\log _{2}\left(1+\xi_{k}^{N L}[n]d_{k}^{-\alpha_{\mathrm{N}}}[n] d_{\bar{k}}^{-\alpha_{\mathrm{L}}}[n]\right)\right.\\&\left.+P_{k}^{\mathrm{N}}[n] P_{\bar{k}}^{\mathrm{N}}[n]\log _{2}\left(1+\xi_{k}^{NN}[n]d_{k}^{-\alpha_{\mathrm{N}}}[n] d_{\bar{k}}^{-\alpha_{\mathrm{N}}}[n]\right)\right)\ge \eta, \tag{19}
\end{align*} 	
where
$\xi_{k}^{L L}[n]= \frac{p_{\bar{k}}|\beta_{0}|^2}{\sigma^2}\left|\left(\mathbf{h}_{k}^{L^{'},(l-1)}[n]\right)^{H} \Theta[n] \mathbf{h}_{\bar{k}}^{L^{'},(l-1)}[n]\right|^{2}$, \\$\xi_{k}^{L N}[n]=\frac{p_{\bar{k}}|\beta_{0}|^2}{\sigma^2} \left|\left(\mathbf{h}_{k}^{L^{'}, (l-1)}[n]\right)^{H} \Theta[n]\mathbf{ \widetilde{h}}_{\bar{k}}^{}\right|^{2}$, \\$\xi_{k}^{N L}[n]=\frac{p_{\bar{k}}|\beta_{0}|^2}{\sigma^2} \left|\left(\mathbf{ \widetilde{h}}_{k}^{}\right)^{H} \Theta[n] \mathbf{h}_{\bar{k}}^{L^{'}, (l-1)}[n]\right|^{2}$,\\ $\xi_{k}^{NN}[n]= \frac{p_{\bar{k}}|\beta_{0}|^2}{\sigma^2}\left|\left(\mathbf{ \widetilde{h}}_{k}^{}\right)^{H} \Theta[n] \mathbf{ \widetilde{h}}_{\bar{k}}^{}\right|^{2}$.}

\textcolor{blue}{To deal with the non-convex constraint (19),  we  introduce  the slack variables $\mathbf{x}={\{x_k[n], \forall k,n\}}$, $\mathbf{y}={\{y_k[n], \forall k,n\}}$, and $\mathbf{z}={\{z_k[n], \forall k,n\}}$ into the rate function (11). Thus,  $	\mathbb{E}\left[R_{k}[n]\right]$ can be rewritten as
\begin{align*}
		\mathbb{E}\left[R_{k}[n]\right]&=\frac{1}{x_{k}[n] x_{\bar{k}}[n]} \log _{2}\left(1+\frac{\xi_{k}^{L L}[n]}{y_{k}^{\alpha_{\mathrm{L}} / 2}[n] y_{\bar{k}}^{\alpha_{\mathrm{L} / 2}}[n]}\right)\\
	&+\frac{1}{x_{k}[n] z_{\bar{k}}[n]} \log _{2}\left(1+\frac{\xi_{k}^{L N}[n]}{y_{k}^{\alpha_{\mathrm{L}} / 2}[n] y_{\bar{k}}^{\alpha_{\mathrm{N} / 2}}[n]}\right)\\
	&+\frac{1}{z_{k}[n] x_{\bar{k}}[n]} \log _{2}\left(1+\frac{\xi_{k}^{N L}[n]}{y_{k}^{\alpha_{\mathrm{N}} / 2}[n] y_{\bar{k}}^{\alpha_{\mathrm{L} / 2}}[n]}\right)\\
	&+\frac{1}{z_{k}[n] z_{\bar{k}}[n]} \log _{2}\left(1+\frac{\xi_{k}^{NN}[n]}{y_{k}^{\alpha_{\mathrm{N}} / 2}[n] y_{\bar{k}}^{\alpha_{\mathrm{N} / 2}}[n]}\right),\tag{20}
\end{align*}
where 
\begin{align*}
	&x_{k}[n] \geq 1+a e^{\left(-b\left[\psi_{k}[n]-a\right]\right)}, \tag{21}\\
%	&z_{k}[n] \geq 1+\frac{1}{a} e^{\left(b\left[\phi_{k}[n]-a\right]\right)},  \tag{26}\\
&y_{k}[n] \geq\left\|\mathbf{q}[n]-\mathbf{w}_{k}\right\|^{2}+h[n]^{2},\tag{22}\\
&z_{k}[n] \geq 1+\frac{1}{a} e^{\left(b\left[\phi_{k}[n]-a\right]\right)}. \tag{23}
\end{align*} 
Furthermore,
\begin{align*}
&\psi_{k}[n] \leq \frac{180}{\pi} \arctan \left(\frac{h[n]}{\left\|\mathbf{q}[n]-\mathbf{w}_{k}\right\|}\right), \tag{24}\\
	 &\phi_{k}[n] \geq \frac{180}{\pi} \arctan \left(\frac{h[n]}{\left\|\mathbf{q}[n]-\mathbf{w}_{k}\right\|}\right), \tag{25}
\end{align*}
are  the relaxed constraints for the sake of handling the non-affine constraints (5). We can prove by contradiction that constraints (21)-(25) must hold with equalities to ensure that the objective value of problem (18) does not decrease.  
Note that, after the variable replacement, $\bar{R}_k[n]$ in (20) is jointly convex with respect with $x_k[n]$, $y_k[n]$, and $z_k[n]$.}

 \textcolor{blue}{Although constraint (24) is non-convex,  the right-hand-side (RHS)
of (24) is convex with respect to $||\mathbf{q}[n]-\mathbf{w}_k||$. Since the first-order Taylor approximation of a convex function is a global underestimator, it can be applied at any
local points $x_k^{(l)}[n]$, $y_{k}^{(l)}[n]$,  $z_{k}^{(l)}[n]$, and $||\mathbf{q}^{(l)}[n]-\mathbf{w}_k||$ in the $l$th
iteration for (20) and (24), i.e., (26) which is shown at the top of the next page, and
\begin{align*}
		&\psi_{k}[n] \leq \frac{180}{\pi} \arctan \left(\frac{h[n]}{\left\|\mathbf{q}[n]-\mathbf{w}_{k}\right\|}\right) \\
		&=\frac{180}{\pi}\left(F_{k}^{(l)}[n]-G_{k}^{(l)}[n]\left(\left\|\mathbf{q}[n]-\mathbf{w}_{k}\right\|-\left\|\mathbf{q}^{(l)}[n]-\mathbf{w}_{k}\right\|\right)\right), \tag{27}
\end{align*}}
\begin{table*}
\begin{footnotesize}
	\textcolor{blue}{\begin{align*}
		&\frac{\log _{2}\left(B^{(l)}[n]\right)}{x^{(l)}_{k}[n] x^{(l)}_{\bar{k}}[n]}+\frac{\log _{2}\left(C^{(l)}[n]\right)}{x^{(l)}_{k}[n] z^{(l)}_{\bar{k}}[n]} +\frac{\log _{2}\left(D^{(l)}[n]\right)}{z^{(l)}_{k}[n] x^{(l)}_{\bar{k}}[n]} +\frac{\log _{2}\left(E^{(l)}[n]\right)}{z^{(l)}_{k}[n] z^{(l)}_{\bar{k}}[n]}  
		-\left(\frac{\log _{2}\left(B^{(l)[n]}\right)}{x_{k}^{(l)}[n]\left(x_{\bar{k}}^{(l)}[n]\right)^{2}}+\frac{\log _{2}\left(D^{(l)[n]}\right)}{z_{k}^{(l)}[n]\left(x_{\bar{k}}^{(l)}[n]\right)^{2}}\right)(x_{\bar{k}}[n]-x_{\bar{k}}^{(l)}[n])
		-\left( \frac{\log _{2}\left(B^{(l)}[n]\right)}{\left(x_{k}^{(l)}[n]\right)^{2} x_{\bar{k}}^{(l)}[n]}\right.\\
		&\left.+ \frac{\log _{2}\left(C^{(l)}[n]\right)}{\left(x_{k}^{(l)}[n]\right)^{2} z_{\bar{k}}^{(l)}[n]} \right)(x_{k}[n]-x_{k}^{(l)}[n])
		-\left(\frac{\log _{2}\left(C^{(l)}[n]\right)}{x_{k}^{(l)}[n]\left(z_{\bar{k}}^{(l)}[n]\right)^{2}}+\frac{\log _{2}\left(E^{(l)}[n]\right)}{z_{k}^{(l)}[n]\left(z_{\bar{k}}^{(l)}[n]\right)^{2}}\right)\left(z_{\bar{k}}[n]-z_{\bar{k}}^{(l)}[n]\right)
		-\left(\frac{\log _{2}\left(D^{(l)}[n]\right)}{x_{\bar{k}}^{(l)}[n]\left(z_{{k}}^{(l)}[n]\right)^{2}}+\frac{\log _{2}\left(E^{(l)}[n]\right)}{z_{\bar{k}}^{(l)}[n]\left(z_{{k}}^{(l)}[n]\right)^{2}}\right)\left(z_{{k}}[n]\right.\\
		&\left.-z_{{k}}^{(l)}[n]\right)
		- \left(\frac{\alpha_L \xi^{LL}_k[n] \left(y_{k}^{(l)}[n]\right)^{{\frac{-\alpha_L-1}{2}}} \log _{2} e}{2 x_{k}^{(l)}[n] x_{\bar{k}}^{(l)}[n]\left(y_{\bar{k}}^{(l)}[n]\right)^{\frac{\alpha_L}{2}} B^{(l)}[n]}
		+ \frac{\alpha_L \xi^{LN}_k[n] \left(y_{k}^{(l)}[n]\right)^{{\frac{-\alpha_L-1}{2}}} \log _{2} e}{2 x_{k}^{(l)}[n] z_{\bar{k}}^{(l)}[n]\left(y_{\bar{k}}^{(l)}[n]\right)^{\frac{\alpha_N}{2}} C^{(l)}[n]}
		+ \frac{\alpha_N \xi^{NL}_k[n] \left(y_{k}^{(l)}[n]\right)^{{\frac{-\alpha_N-1}{2}}} \log _{2} e}{2 z_{k}^{(l)}[n] x_{\bar{k}}^{(l)}[n]\left(y_{\bar{k}}^{(l)}[n]\right)^{\frac{\alpha_L}{2}} D^{(l)}[n]}\right.\\
		&\left.+ \frac{\alpha_N \xi^{NN}_k[n] \left(y_{k}^{(l)}[n]\right)^{{\frac{-\alpha_N-1}{2}}} \log _{2} e}{2 z_{k}^{(l)}[n] z_{\bar{k}}^{(l)}[n]\left(y_{\bar{k}}^{(l)}[n]\right)^{\frac{\alpha_N}{2}} E^{(l)}[n]}\right) (y_{k}[n]-y_{k}^{(l)}[n])
		- \left(\frac{\alpha_L \xi^{LL}_k[n] \left(y_{\bar{k}}^{(l)}[n]\right)^{\frac{-\alpha_L-1}{2}} \log _{2} e}{2 x_{k}^{(l)}[n] x_{\bar{k}}^{(l)}[n]\left(y_{k}^{(l)}[n]\right)^{\frac{\alpha_L}{2}} B^{(l)}[n]}
		+ \frac{\alpha_N \xi^{LN}_k[n] \left(y_{\bar{k}}^{(l)}[n]\right)^{\frac{-\alpha_N-1}{2}} \log _{2} e}{2 x_{k}^{(l)}[n] z_{\bar{k}}^{(l)}[n]\left(y_{k}^{(l)}[n]\right)^{\frac{\alpha_L}{2}} C^{(l)}[n]}\right.\\
		&\left.+ \frac{\alpha_L \xi^{NL}_k[n] \left(y_{\bar{k}}^{(l)}[n]\right)^{\frac{-\alpha_L-1}{2}} \log _{2} e}{2 z_{k}^{(l)}[n] x_{\bar{k}}^{(l)}[n]\left(y_{k}^{(l)}[n]\right)^{\frac{\alpha_N}{2}} D^{(l)}[n]}
		+ \frac{\alpha_N \xi^{NN}_k[n] \left(y_{\bar{k}}^{(l)}[n]\right)^{\frac{-\alpha_N-1}{2}} \log _{2} e}{2 z_{k}^{(l)}[n] z_{\bar{k}}^{(l)}[n]\left(y_{k}^{(l)}[n]\right)^{\frac{\alpha_N}{2}} E^{(l)}[n]}\right)
 (y_{\bar{k}}[n]-y^{(1)}_{\bar{k}}[n]), \tag{26}\\
	\end{align*}}
\end{footnotesize}\vspace{-0.7cm}\hrule \vspace{-0.6cm}\end{table*}\textcolor{blue}{where $B^{(l)}[n]=1+\frac{\xi^{LL}_k[n]}{\left(y_{k}^{(l)}[n]\right)^{\alpha_L / 2}\left(y_{\bar{k}}^{(l)}[n]\right)^{\alpha_L / 2}}$, $C^{(l)}[n]=1+\frac{\xi^{LN}_k[n]}{\left(y_{k}^{(l)}[n]\right)^{\alpha_L / 2}\left(y_{\bar{k}}^{(l)}[n]\right)^{\alpha_N / 2}}$, $D^{(l)}[n]=1+\frac{\xi^{NL}_k[n]}{\left(y_{k}^{(l)}[n]\right)^{\alpha_N / 2}\left(y_{\bar{k}}^{(l)}[n]\right)^{\alpha_L / 2}}$, and  $E^{(l)}[n]=1+\frac{\xi^{NN}_k[n]}{\left(y_{k}^{(l)}[n]\right)^{\alpha_N / 2}\left(y_{\bar{k}}^{(l)}[n]\right)^{\alpha_N / 2}}$.  Furthermore, $F_{k}^{(l)}[n]=\arctan \left(\frac{h[n]}{\left\|\mathbf{q}^{(l)}[n]-\mathrm{w}_{k}\right\|}\right)$ and $G_{k}^{(l)}[n]=\frac{h[n]}{\left\|\mathbf{q}^{(l)}[n]-w_{k}\right\|^{2}+H^2}.$}\\

\textcolor{blue}{With (26)-(27), problem (18) can be reformulated into the following convex optimization problem,
\begin{align*}
	&\max _{\mathbf{Q},  \Psi_{k}, \Phi_{k} \mathbf{x}_{k}, \mathbf{z}_{k}, \mathbf{y}_{k}, \eta} \eta \tag{28}\\
	&\text { s.t. } (1)-(2),(21)-(23),  (25)-(27),
\end{align*}
where  $\Psi_{k}=\left\{\psi_{k}[n], \forall k, n\right\}$ and  $\Phi_{k}=\left\{\phi_{k}[n], \forall k, n\right\}$.  As such, problem  (28) can be efficiently solved by $\mathrm{CVX}$ \cite{cvx}.}
%\begin{footnotesize}
%%		\begin{algorithm}[t]
%%	\caption{Proposed algorithm for problem (15)} %算法的名字11
%%	\begin{algorithmic}[1]
%%		\STATE $\mathbf{Initialization:}$Initialize  $\mathbf{A}^{(0)}$, $\mathbf{\Theta}^{(0)}$, $\mathbf{Q}^{(0)}$ and
%%		iteration number $l = 0$. Set $\bar{R}^{(0)}_k$ and $\bar{R}^{(0)}_{\bar{k}}$ by using (14) with given ($\mathbf{A}^{(0)}$, $\mathbf{\Theta}^{(0)}$, $\mathbf{Q}^{(0)}$).
%%		\REPEAT
%%			\STATE Solve problem (16) for given ($\mathbf{\Theta}^{(l)}$, $\mathbf{Q}^{(l)}$), and denote the optimal solution as ($\mathbf{A}^{l+1}$).
%%			\STATE Solve problem (17) for given ($\mathbf{A}^{(l)}$, $\mathbf{Q}^{(l)}$), and denote the optimal solution as ($\mathbf{\Theta}^{l+1}$).
%%				\STATE Solve problem (32) for given ($\mathbf{\Theta}^{(l)}$, $\mathbf{A}^{(l)}$), and denote the optimal solution as ($\mathbf{Q}^{l+1}$).
%%			\STATE With given ($\mathbf{A}^{(l+1)}$, $\mathbf{\Theta}^{(l+1)}$, $\mathbf{Q}^{(l+1)}$), update $\bar{R}^{(l+1)}_k$  and $\bar{R}^{(l+1)}_{\bar{k}}$ by using (14).
%%				\STATE Update $l=l+1$.
%%		\UNTIL { $\frac{\text{min}(\bar{R}^{(l+1)}_k, \bar{R}^{(l+1)}_{\bar{k}})-\text{min}(\bar{R}^{(l)}_k, \bar{R}^{(l)}_{\bar{k}})}{\text{min}(\bar{R}^{(l)}_k, \bar{R}^{(l)}_{\bar{k}})}<\epsilon$.}
%%\end{algorithmic}
%%\end{algorithm}
%\end{footnotesize}
	\subsubsection{UAV Vertical Trajectory Optimization}With the optimal $\mathbf{A}$, $\mathbf{Q}$, and $\mathbf{\Theta}$ obtained by problem (12), (17), and (28), the UAV vertical trajectory optimization problem can be reformulated as
	\begin{align*}
		&\max _{\mathbf{H},  \mathbf{\Psi_k}, \eta} \eta \tag{29}\\
		&\text { s.t. } (3), (5), (11b),
	\end{align*}
\quad 
	Since problems (18) and (29) are similar in form and differ only slightly in terms of optimization variables $\mathbf{H}$,  the procedure for solving problem (18) can be similarly applied  to solve problem (29). We omit the detailed derivation owing to the page limitation.
\textcolor{blue}{\subsubsection{Overall Algorithm}
  By applying our proposed algorithm, problem (15) can be  solved by alternately optimizing variables $\mathbf{A}$, $\mathbf{\Theta}$,  $\mathbf{Q}$, and $\mathbf{H}$, while its solution converges to a preset accuracy $\epsilon$.  Note that the binary solution can be reconstructed with high precision from the obtained continuous variables of GN’s transmission scheduling by applying the proposed reconstruction method in \cite{3}.
Since the four subproblems are solved by applying CVX via the standard interior point method, their computational complexity can be obtained as $O_{1}\left((KN)^{3.5}\text{log}(1/\epsilon)\right)$, 
$O_{2}\left(\sqrt{M}\text{log}(1/\epsilon)\left((KN+1)M^{3}+(K\right.\right.\\\left.\left.N+1)^2M^{2}+(KN+1)^3\right)\right)$, $O_{3}\left((2N+6KN)^{3.5}\text{log}(1/\epsilon)\right)$, and $O_{4}\left((2N+6KN)^{3.5}\text{log}(1/\epsilon)\right)$, respectively. Thus, the total computational complexity of our proposed algorithm is in the order of $O_1+O_2+O_3+O_4$.} \\
%		\begin{figure}
%		\centering
%		\includegraphics[width=80mm]{l1.pdf}\\
%		\caption{UAV trajectories by different elements. }
%		\label{fig:env}
%	\end{figure} 
%	
%	
%	\begin{figure}
%		\centering
%%		\includegraphics[width=80mm]{lyf2.png}\\
%		\caption{Average rate by different elements verus
%			T. }
%		\label{fig:env}
%	\end{figure}
\begin{figure*}[htbp] %通栏
	\begin{minipage}[t]{0.33\linewidth} %调节两个子图左右间距
		\centering
		\includegraphics[width=2.2in, height=1.6in]{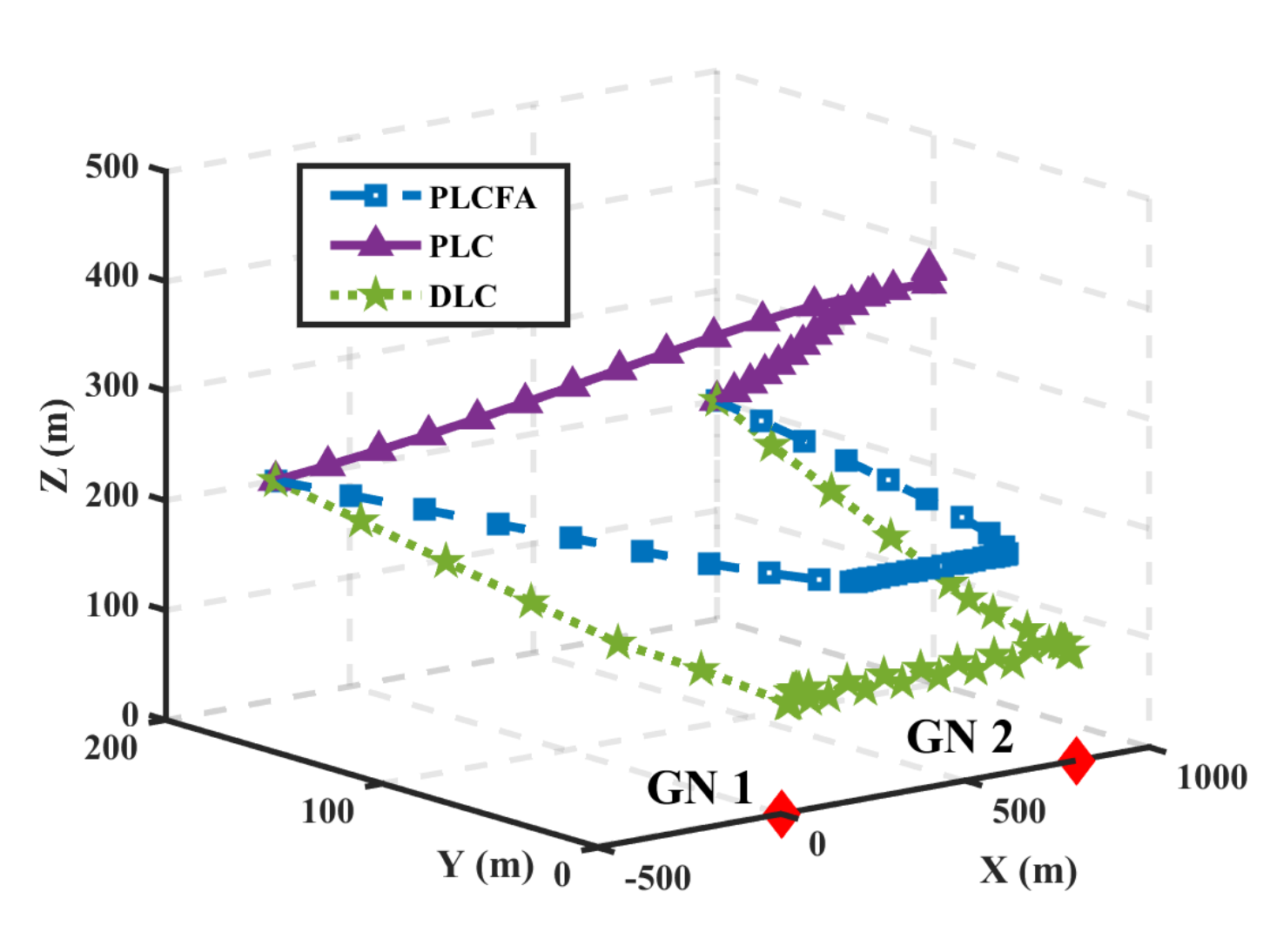} %调节单个子图大小
		\caption{\textcolor{blue}{UAV’s trajectories.}} %子图下标题
		\label{fig:side:b} %引用标签
	\end{minipage}%
	\begin{minipage}[t]{0.33\linewidth}
		\centering
		\includegraphics[width=2.2in, height=1.6in]{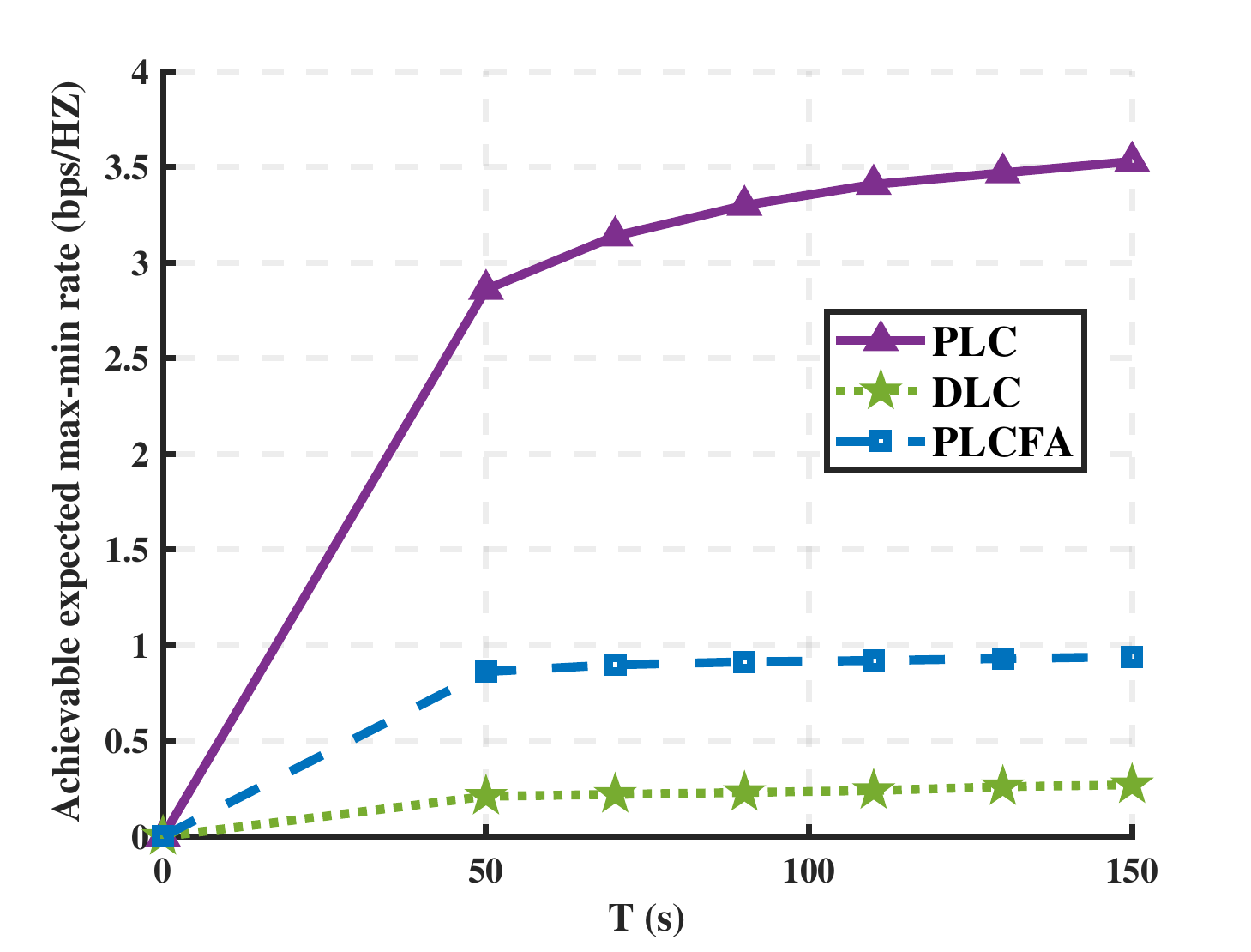}
		\caption{\textcolor{blue}{Achieved expected max-min  \\rate  versus $T$.}}
		\label{fig:side:c}
	\end{minipage}%
	\begin{minipage}[t]{0.33\linewidth}
		\centering
		\includegraphics[width=2.2in, height=1.6in]{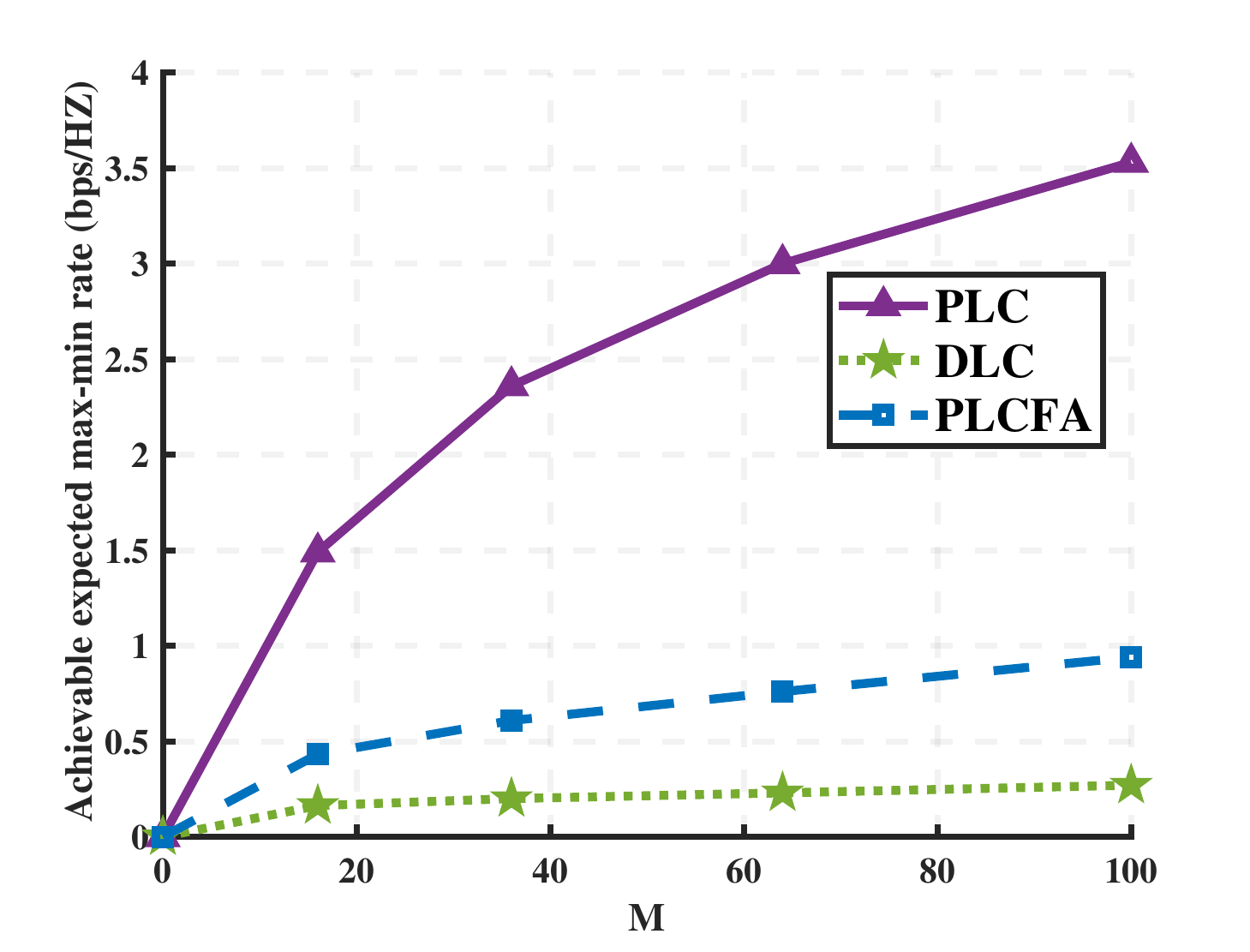}
		\caption{\textcolor{blue}{Achieved expected max-min\\ rate  versus $M$.}}
		\label{fig:side:d}
	\end{minipage}\vspace{-0.6cm}
\end{figure*} 
	\section{Simulation Results}
In this section, we provide  simulation results to show the interesting elevation angle and distance trade-off  of the 3D UAV trajectory design under the PLC model (denoted as PLC for brevity). The following schemes are used for comparison: 1) UAV horizontal trajectory design under the PLC scheme with fixed altitude being $H=200$ m (denoted as PLCFA); 2) 3D UAV trajectory design under the  LC model (denoted as DLC). We assume that GN1 and GN 2 are located at (0, 0, 0) m and (800, 0, 0) m, respectively. \textcolor{blue}{Unless otherwise stated, the simulation results are set as:  $\hat{v}_{max}=40$ m/s, $\tilde{v}_{max}=20$ m/s  $p_{1}=p_2=0.1 \mathrm{~W}$, $\beta_{0}=-40$  $\mathrm{dB}$, $H_{\rm {min}}=100$ m, $H_{\rm {max}}=500$ m, $d=\frac{\lambda}{2}$, $\sigma^{2}=-169$ $\mathrm{dBm}$, $\alpha_{\mathrm{L}}=2.2$, $\alpha_{\mathrm{N}}=3.2$, $a=11.95$, and $b=0.14$ \cite{7,6}. Furthermore,  the initialized trajectory of the UAV is set to be a straight-line trajectory with a fix altitude $H = 200$ m, i.e., the UAV flies from $\mathbf{q}_0= [-200, -200]$ to  $\mathbf{q}_{F}=[1000, 200]$ at its maximum flying speed.}

Fig. 2 shows the different 3D UAV trajectories by three schemes   with $T=150$ s. It can be seen that the UAV in the DLC  first descends quickly to $H_{\rm {min}}$, hovers over GN 1, then flies horizontally at its maximum speed, and hovers over GN 2. Finally, the UAV rises and flies back to the final location. This is because hovering above each GN at $H_{\rm {min}}$ suffers from the least path loss for the cascade channel between the UAV and the GNs.   By contrast, in the PLC scheme, the UAV first ascends rapidly to increase the elevation angle between the UAV and each GN for a higher LoS probability. Then, the UAV hovers above the midpoint of the two GNs, which results in less path loss while maintaining a larger LoS probability of the cascaded channel. Compared to the PLC scheme that the UAV hovers only above the midpoints of the two GNs, the UAV in the PLCFA scheme hovers close to both sides of the midpoint of the two GNs. This is because that although hovering at the midpoint of the two GNs can maximize the LoS probability of the cascaded channel, i.e., $P_{k}^{\mathrm{L}}[n] P_{\bar{k}}^{\mathrm{L}}[n]$,  the larger path loss of the communication links also degrades the rate performance. Therefore, our proposed PLC scheme can take advantage of the additional design brought by the UAV vertical trajectory to obtain a more efficient angle-distance trade-off than the PLCFA scheme. 

Fig. 3 illustrates the  achieved expected max-min  rates of different schemes versus $T$ when $M=100$. It can be seen that the PLCFA scheme achieves larger rates than the DLC scheme, which indicates the necessity of adopting the more accurate PLC model to describe the LoS/NLoS channel states in the UAV-borne RIS communication system. Furthermore, our proposed PLC scheme significantly improves performance over the PLCFA scheme.  The reason is that the additionally designing the UAV vertical trajectory can further increase the LoS probability  of the cascaded channel.  Fig. 4 shows the achieved expected max-min  rates for different schemes with $T = 150$ s versus different $M$. As expected, the rate performance is significantly increased when more elements are equipped in the RIS due to the larger passive beamforming gain. In practice, we cannot increase $M$ indefinitely to obtain higher rates due to the limitation of the size of rotary-wing UAVs. An oversized $M$ will result in a larger RIS size, which will increase the weight of the UAV and result in greater energy consumption (i.e., shorter endurance). Therefore, an interesting trade-off exists between $M$ (i.e., related to the RIS size) and the UAV energy consumption for rate improvement.
\section{Conclusion}
	This paper investigated the potential of rate enhancement for the aerial RIS-aided communication system under  the accurate PLC model in the dense urban environment. The objective was to maximize the minimum average achievable rate.  We proposed an efficient iterative algorithm to jointly  optimize the communication schedule, the  RIS's phase shift, and the 3D UAV trajectory.  Numerical results showed that the proposed scheme has a significant improvement compared to the conventional DLC scheme. Furthermore, our proposed scheme enjoys the additional gain of elevation angle-dependent 3D UAV trajectory design and can effectively balance the elevation angle and distance trade-off between the UAV and the GNs, whereas the DLC scheme cannot. This validates the practical importance of considering the more accurate PLC model to support UAV-borne RIS communications in urban environments.
	
\textcolor{blue}{ Note that in the suburban/urban/dense urban environment, the error in the approximated PLC model \cite{R1} is extremely small and thus can be ignored. However, this PLC model suffers from errors in the high-rise urban environment, which may cause performance loss. In this case, the hybrid \emph{offline-online} 3D UAV trajectory design proposed in [17] may be an alternative to better characterize the real-time location-dependent air-ground channel states, which is interesting to study whether the performance loss can be further compensated by employing additional online design.}

	\ifCLASSOPTIONcaptionsoff
	\newpage
	\fi

	
	
	%\bibliographystyle{IEEEtran}
	%\textit{\begin{thebibliography}{1}
	%  S. Chen, Y. -C. Liang, S. Sun, S. Kang, W. Cheng and M. Peng, "Vision, Requirements, and Technology Trend of 6G: How to Tackle the Challenges of System Coverage, Capacity, User Data-Rate and Movement Speed," in IEEE Wireless Communications, vol. 27, no. 2, pp. 218-228, April 2020, doi: 10.1109/MWC.001.1900333.
	%
	%\end{thebibliography}
	%}
	\bibliographystyle{IEEEtran}
	\bibliography{mycite}
\end{document}